\PassOptionsToPackage{svgnames}{xcolor}
\documentclass[11pt]{article}

\usepackage{amsmath,amssymb,amsfonts,amsthm,setspace,tikz,natbib,dsfont, xcolor, array,colortbl,mathrsfs,pifont,wasysym}
\usetikzlibrary{calc}
\usetikzlibrary{patterns}
\usetikzlibrary{shapes.geometric}
\usetikzlibrary{positioning}
\usetikzlibrary{trees}
\usepackage{centernot}
\usepackage{xspace}
\usepackage{mathtools}
\usepackage[scr=boondox,scrscaled=1.05]{mathalfa}
\usepackage{multirow}
\usepackage{graphicx}
\usepackage{subcaption}
\usepackage{bm}
\usepackage{accents} 

\usepackage{thmtools,thm-restate}
\usepackage{enumitem}

\usepackage[makeroom]{cancel}

\usepackage[letterpaper]{geometry}
\definecolor{lamx}{HTML}{CB0404}
\definecolor{lamy}{HTML}{732255}
\definecolor{lama}{HTML}{309898}
\definecolor{lam4}{HTML}{e07c02}
\definecolor{lam5}{HTML}{B2C6D5}

\usepackage[colorlinks=true,urlcolor=lamy,linkcolor=lamy,citecolor=lamy]{hyperref}

\onehalfspace

\newtheoremstyle{mytheoremstyle} 
    {\topsep}                    
    {\topsep}                    
    {\itshape}                   
    {}                           
    {\sc}                   
    {.}                          
    {.5em}                       
    {}  

\theoremstyle{mytheoremstyle}

\newtheorem{definition}{Definition}
\newtheorem{assumption}{Assumption}
\newtheorem{theorem}{Theorem}

\newtheorem{corollary}{Corollary}

\newtheoremstyle{scfont} 
    {\topsep}                    
    {\topsep}                    
    {}                   
    {}                           
    {\scshape}                   
    {}                          
    {.5em}                       
    {\textbf{Axiom \thmnumber{#2}\thmname{#1}}\thmnote{---{\textsc{#3}.}}}

\theoremstyle{scfont}

\theoremstyle{remark}

\newenvironment{tproof}[1]{\noindent \emph{Proof of Theorem \ref{#1}. \phantomsection\label{pf:#1}\addcontentsline{toc}{subsection}{Pf. Thm. \ref{#1}}}}{\hfill$\blacksquare$ \\}

\newcounter{ex}

\usepackage{titlesec}

\titleformat*{\section}{\large\bf\centering}

\titleformat{\subsection}[runin]
  {\bfseries}         
  {\thesubsection \quad }                  
  {0pt}               
  {}[.]               

\titlespacing*{\section}{0pt}{3.5ex plus 1ex minus .2ex}{2.5ex plus .2ex}

\renewcommand{\emptyset}{\varnothing}
\renewcommand{\phi}{\varphi}
\renewcommand{\implies}{\rightarrow}

\def \R{\mathbb{R}}

\def \A{\mathcal{A}}
\def \F{F}
\def \H{\mathcal{H}}
\def \W{\Omega}
\def \w{\omega}

\def\ss{\sigma}

\def \BR{\textup{BR}}
\def \BRR{\textup{BR}^{fr}}
\def \IC{\textup{IC}}

\def \<{\langle}
\def \>{\rangle}

\def \x{^{\star}}

\def \1{\mathbf{1}}
\def \0{\mathbf{0}}

\def\Rev{\mathcal{R}}
\def\Qev{\mathcal{Q}}

\def\xx{\textcolor{lamx}{\texttt{x}}}
\def\yy{\textcolor{lamy}{\texttt{y}}}
\def\aa{\textcolor{lama}{\texttt{a}}}
\def\bb{\textcolor{lama}{\texttt{b}}}

\newcommand{\df}[1]{\textup{\textbf{#1}}}

\DeclareMathSymbol{\shortminus}{\mathbin}{AMSa}{"39}

\newcommand{\scref}[1]{\emph{\ref{#1}}}

\DeclareMathOperator*{\argmin}{arg\,min}
\DeclareMathOperator*{\argmax}{arg\,max}

\newcommand{\hide}[1]
{
}

\title{Iterated Revelation: \\
\large How to Incentivize Experts to Reveal Novel Actions\thanks{
  I would like to thank the seminar and conference audiences at Paris-Sorbonne, Pompeu Fabra, Universitat Autònoma de Barcelona, NYU Shanghai, University of Hong Kong, Hong Kong University of Science and Technology, University College London, SAET 2023, TUS-X, Aarhus Workshop on Known and Unknown Unknowns, and particular thanks to David Levine, Alexey Kushnir, Philip Neary, Paula Onuchic, Antonio Penta, Burkhard Schipper, Alex Smolin, John Quiggin,  David Walker-Jones, and Joel Watson for their conversations and feedback. 
}}

\author{
Evan Piermont\footnote{Royal Holloway, University of London, Department of Economics; \href{mailto:evan.piermont@rhul.ac.uk}{\tt evan.piermont@rhul.ac.uk}.}
} 

\begin{document}

\maketitle

\vspace{1cm}

\begin{abstract}
\footnotesize
\noindent I examine how a decision maker can incentivize an expert to reveal novel actions, expanding the set from which he can choose, without making ex-ante commitments regarding as-of-yet unrevealed actions. The outcomes achievable by any (incentive compatible) mechanism are characterized by the \emph{iterated revelation protocol}: a simple dynamic interaction where, each round, the expert reveals novel actions and the decision maker adds actions to a shortlist; when nothing novel is revealed, the mechanism ends with the \emph{expert} choosing an action from the shortlist. Greedy strategies---where the decision maker optimizes myopically---delineate the decision maker's maximal payoff achievable by any efficient mechanism.
\vspace{1cm}

\noindent\textsc{Keywords}: Iterated Revelation, Project Choice, Incomplete Contracts; Antitrust Regulation\\
\textsc{JEL Classification}: D81, D83, D86
\end{abstract}


\newpage

\section{Introduction}
\label{sec:intro}

\setlength{\abovedisplayskip}{5pt}
\setlength{\belowdisplayskip}{5pt}

A decision maker (he) must choose a project and holds the ultimate authority on which project gets enacted.
In complex environments, where he will not in general be able to conceive of all relevant courses of action, he might enlist the help of an expert (she) who can reveal novel projects.
When such revelations are voluntary, the expert, who has a vested interest in the outcome, may strategically decide not to reveal the full set of projects she is aware of. This paper asks how the decision maker might incentivize full revelation and ensure efficient outcomes, in particular, when he is unable to commit himself to a specific choice conditional on the expert's revelation. This inability to commit arises naturally, since before the expert's revelation, the decision maker may not be able to express the details of the problem necessary to describe what he would eventually choose.

An economically important application of this analysis is to antitrust regulation \citep{besanko1993contested,lyons2003could,nocke2013merger}. Here, a firm (the expert) proposes potential mergers to a regulator (the decision maker). The regulator is tasked with approving or rejecting mergers, but will not generally be able to conceive of all potential merger opportunities. The main results of this paper characterize the regulator's optimal efficient mechanism for approving mergers, and significantly, show that such a mechanism can be implemented without the need to ex-ante commit to (or describe) which mergers it would approve.

As in the literature on project choice---e.g., \cite{armstrong2010model} and \cite{guo2023regret}---each project yields a known payoff to both the decision maker and the expert: the only uncertainty in the baseline model regards the universe of potential projects.\footnote{Section \ref{sec:payoff_uncertainty} discusses a generalization of the model with payoff uncertainty and discusses the extent to which the results of this paper can be generalized.} Some projects cannot be chosen by the decision maker unless the expert reveals them. Further, the decision maker does not know the full set of projects that can be revealed by the expert. As such, in order to obtain desirable outcomes, the decision maker must construct an incentive scheme (i.e., mechanism) to incentivize the expert to reveal.

This paper diverges from the existing literature on project choice in two important ways. First, I consider the case where the decision maker is unable to make ex-ante commitments about how he will select from as-of-yet unrevealed projects. 
This could be because he is unaware of novel projects, does not possess the language to articulate them, or because of technological or legal restrictions on commitment. This assumption precludes the use of direct mechanisms, of interest in the literature on project choice, or welfare standards, ubiquitous in the analysis of antitrust regulation \citep{farrell2006economics}. 

Second, I focus on the role of informational spillover in revelations. In general, it may be impossible to reveal some projects without revealing others. For example: A firm entertains two potential targets for acquisition that the regulator has not considered; it could profitably acquire either or both of the targets. If the firm proposes to acquire both targets, it inadvertently reveals the possibility of merging with each target individually as well. In this paper, I allow for arbitrary information spillover and examine how this novel channel influences the rents the expert can procure from the decision maker.


To address how to incentivize revelation and ensure efficient outcomes without the ability to commit to projects before they are revealed (i.e., without direct mechanisms), I introduce the \emph{Iterated Revelation Protocol} (IRP). The IRP is a straightforward dynamic interaction where, in each round, the expert can (voluntarily) expand the decision maker's choice set by revealing novel projects, and the decision maker can add a new project to a `shortlist.' This back and forth ends when nothing novel is revealed by the expert.  The project that gets enacted is then chosen from the shortlist \emph{by the expert}. 

 The critical feature of the IRP facilitating full revelation is that once a project is added to the shortlist it cannot be removed, and so, the expert is forever afforded the possibility of choosing it. Implicitly, this ensures that no matter how the interaction unfolds after further revelation, the expert's payoff is bounded below by her most preferred project already on the shortlist, and therefore, cannot decline. Moreover, the IRP is practical because the description of the interaction---hence the ability to commit to follow it---does not depend on what has not yet been revealed. In other words, while the outcome of the process may be inexpressible, the process itself is.


Theorem \ref{thm:mondisclosure} shows that the outcome of \emph{any} incentive compatible mechanism can be implemented as a best response to some strategy of the decision maker in the IRP. 
 As such, the IRP can---like direct mechanisms---serve as a universal tool to characterize which outcomes are implementable, but---unlike direct mechanisms--- it does not require the ex-ante ability to express every potential action. Theorem \ref{thm:full_implements} then characterizes the subset of these outcomes which can be fully implemented---those that are the \emph{unique} outcome of all best responses. 

I then use the universality of the IRP as a tool to study which projects can be implemented efficiently. In Section \ref{sec:eff}, I examine the case where the decision maker must select Pareto efficient projects. This is a natural restriction in many applications. For example, even when the regulator cares principally about consumer welfare, it may be politically untenable to implement a merger while rejecting a different merger that was better for the firm and no worse for consumers.  Under the requirement of efficiency, the decision maker can do no better than to use a  \emph{greedy strategy}: in each round, he shortlists his most preferred project, subject to the implicit incentive constraint that the expert prefers it to all prior shortlisted projects. Specifically, Theorem \ref{thm:recvbestMech} shows playing a greedy strategy in the IRP implements the decision maker's preferred project across those implemented by \emph{any} efficient and incentive compatible mechanism, \emph{no matter what} the expert can express.

The IRP can be thought of as a particular process by which the decision maker and expert negotiate how to split the surplus created by their cooperation. Greedy strategies thus trace out the maximal surplus that can be captured by the decision maker under any bargaining protocol. Applying this result to the case of antitrust regulation yields the following insight: Rather than committing to rules that attempt to anticipate future merger proposals and specify conditional approval ex-ante, the regulator is always better off committing to a particular negotiation \emph{process}, the IRP, and then maximizing greedily.

While arbitrary mechanisms can be simulated within the IRP via an appropriate strategy (Theorem \ref{thm:mondisclosure}), identifying such a strategy may require a level of forward planning beyond the cognitive grasp of the decision maker. This is especially the case under bounded awareness or with imprecise beliefs about what has yet to be revealed. It is particularly remarkable, then, that when choosing efficient outcomes, forward planning is thoroughly unnecessary as greedy strategies are entirely backwards looking and computationally trivial to execute.



In Section \ref{sec:general_strategic_analysis}, I consider a more general strategic analysis, dropping the requirement of efficiency. The decision maker's strategic concerns when adding a project to the shortlist are captured by the trade-off between (i) maximizing his payoff should the protocol end and the project get chosen, and (ii) minimizing the expert's payoff so as to relax future incentive constraints as much as possible.
Theorem \ref{thm:locally_rat_is_undominated} shows that \emph{locally rational strategies}---those that always shortlist projects at the frontier of this trade-off---correspond to undominated direct mechanisms.  In other words, the decision maker can justify any shortlist proposal so long as no other project has been revealed that is simultaneously better for himself and worse for the expert. Further, Theorem \ref{thm:greedy_is_cautious} shows that \emph{greedy} strategies---those that solve this trade off by maximizing (i) and ignoring (ii)---correspond to cautious direct mechanisms, those that maximize the worst case outcome going forward.

These two results stand as bounds on the strategic predictions of the model. Locally rational strategies are those that optimize with respect to a specific ex-ante belief about the likelihood of what the expert can express. While such strategies are optimal in expectation with respect to said ex-ante belief, they might fare poorly  under other distributions of the expert's type. Thus, if the decision maker cannot formulate a prior, she may want to hedge by acting cautiously. Greedy strategies are maximally cautious, optimizing the worst case outcome across all possible types of the expert. 

$$
\ast \ast \ast 
$$

\medskip

The remainder of this paper takes the following structure: This section concludes with a numerical example that exposes the mechanics of the iterated revelation protocol in the context of antitrust regulation.
Section \ref{sec:model} presents the environment of the model and the IRP, and contains the universality results relating strategies to direct mechanisms.  Section \ref{sec:eff} analyzes the case where the decision maker is constrained to enact efficient outcomes and Section \ref{sec:general_strategic_analysis} considers the case without this constraint.
Section \ref{sec:payoff_uncertainty} examines a generalization of the model where the decision maker faces payoff uncertainty. Finally, a discussion, including relevant literature, is presented in Section \ref{sec:lit}. All proofs are found in the appendix. 

\subsection{Example and Illustration of Results}
\label{sec:ex}

A firm (the expert) can propose specific mergers to an antitrust regulator (the decision maker). The regulator holds the ultimate authority in allowing any merger to proceed but cannot, in general, envision all potential mergers. The firm, which better understands the organization and industry structure, is aware of mergers that are unforeseen by the regulator. 

Each merger $z$ can be described by a point $(z_c, z_f) \in \R^2$, where $z_c$ represents the benefit to the consumers and $z_f$ the benefit to the firm. The firm cares only about maximizing $z_f$, whereas the regulator wants to maximize $z_c$. However, in addition, the regulator is required to choose efficiently: that is, the regulator cannot approve a merger while rejecting some Pareto dominating merger. For example, if the firm makes the regulator aware of  $(z_c, z_f)$ and $(z_c, z'_c)$ for $z_f > z'_f$ then, although the regulator is indifferent between the mergers, it cannot allow the latter while rejecting the former.

The theoretical literature on antitrust regulation (and more broadly, that on project choice) has analyzed this situation through \emph{selection rules} \citep{nocke2013merger,armstrong2010model,guo2023regret}. A selection rule is an ex-ante commitment by the regulator about how it will select a merger from those revealed by the firm.\footnote{In some models the firm can only propose a single merger, so a selection rule is an approval set specifying which mergers would be approved. This is closely related to the literature on delegation \`a la \cite{holmstrom1980theory}.} For example, two widely considered selection rules are the \emph{total welfare standard} and the \emph{consumer welfare standard}, which select from the revealed mergers those which maximize $z_c + z_f$ and $z_c$, respectively.\footnote{Where the interpretation of $z_c$ and $z_f$ is of consumer and producer surplus, respectively. See \cite{farrell2006economics} for a discussion of welfare standards.}

When the firm can strategically decide which mergers to reveal, such simple rules will not in general be optimal for the regulator. However, even beyond optimality, committing to \emph{any} selection rule may be infeasible as it requires commitments regarding mergers that have not yet been revealed. For example, the welfare standards above could only be implemented when $z_c$ and $z_f$ represent specific, ex-ante contractible outcomes; such rules become infeasible in more complex situations where each agent's value of a merger arises as the culmination of many nebulous features, when utility is non-transferable / ordinal, etc.\footnote{In general, the utility values in the paper can be thought of as transmitting only \emph{ordinal} data, further impeding the ability contract on them. See Footnote \ref{ft:ordinal}.} 

\begin{figure}
\centering
\begin{subfigure}[t]{0.47\textwidth}
\centering
\begin{tikzpicture}[scale=.7]
    \foreach \x in {0,1,2,3,4} \draw[gray, thin, dotted] (\x,-1) -- (\x,5);
    \foreach \y in {0,1,2,3,4} \draw[gray, thin, dotted] (-1,\y) -- (5,\y);
    \draw[dotted,thick,gray!30] (4.12,0) arc (-.2:26:4.5);
    \draw[dotted,thick] (4.12,0) arc (-.2:86:4.5);

    \draw[thick,->] (-1,0) -- (5,0) node[anchor=north] {$z_c$};
    \draw[thick,->] (0,-1) -- (0,5) node[anchor=east] {$z_f$};

    \fill[lamy](0,0) circle (5pt) node[anchor=south west] {\yy};
    \fill[lamx] (2,2) circle (5pt) node[anchor=south west] {\xx};
    \fill[lama] (3,3) circle (5pt) node[anchor=south west] {\aa};
    \fill[lama](4,1) circle (5pt) node[anchor=south west] {\bb};
\end{tikzpicture}
\caption{The set of all mergers. Mergers of the same color cannot be revealed independently.}
\end{subfigure}
\hfill
\begin{subfigure}[t]{0.47\textwidth}
\centering
\begin{tikzpicture}[scale=.7]
    \foreach \x in {0,1,2,3,4} \draw[gray, thin, dotted] (\x,-1) -- (\x,5);
    \foreach \y in {0,1,2,3,4} \draw[gray, thin, dotted] (-1,\y) -- (5,\y);
    \draw[dotted,thick,gray!30] (4.12,0) arc (-.2:26:4.5);
    \draw[dotted,thick] (4.12,0) arc (-.2:86:4.5);

    \draw[thick,->] (-1,0) -- (5,0) node[anchor=north] {$z_c$};
    \draw[thick,->] (0,-1) -- (0,5) node[anchor=east] {$z_f$};

    \fill[lamx] (2,2) circle (5pt) node[anchor=south west] {\xx};
    \fill[lama] (3,3) circle (5pt) node[anchor=south west] {\aa};
    \fill[pattern=north east lines, pattern color=gray!50] (-1,-1) rectangle (5,2);
    \fill[gray!30](0,0) circle (3pt) node[anchor=south west] {\texttt{y}};
    \fill[gray!30](4,1) circle (5pt) node[anchor=south west] {\texttt{b}};
\end{tikzpicture}
\caption{The grayed out mergers will never be enacted by the firm, given that $\xx$ is shortlisted.}
\end{subfigure}
\caption{Illustration of possible and feasible mergers.}
\label{fig:mergers}
\end{figure}

The results of this paper show that the IRP simultaneously overcomes both (i) the complexity in solving for the regulator's optimal behavior and (ii) the regulator's inability to commit to a selection rule. This section illustrates this via the following example: The regulator is initially aware of two mergers:
    $$
    \xx = (2,2) \qquad \yy = (0,0)
    $$
The firm is also aware of:
    $$
    \aa = (3,3) \qquad \bb = (4,1)
    $$
To understand the role of information spillover, assume the mergers $\aa$ and $\bb$ are predicated on the same institutional details, so that one cannot be revealed without also revealing the other. These are shown in the left panel of Figure \ref{fig:mergers}.

In the absence of any commitments, the firm would ideally reveal only $\aa$, which is its most preferred of all possible mergers. However, because of information spillover, this is not possible; revealing $\aa$ would also reveal $\bb$, which the regulator prefers but is worse than $\xx$ for the firm. Anticipating that the regulator would then choose $\bb$, the firm will not reveal, leading to an inefficient outcome. 

Now, consider what happens in the IRM. The regulator shortlists $\xx$, which becomes the outside option should no novel merger be agreed upon. With $\xx$ shortlisted, the firm is now secure in revealing $\{\aa,\bb\}$, since if the regulator subsequently shortlists $\bb$, the firm can revert to $\xx$; see the right panel of Figure \ref{fig:mergers}. Anticipating this, the regulator shortlists $\aa$, and the firm enacts it: this is an efficient outcome.\footnote{This particular sequence is represented within dynamic game in Figure \ref{fig:game-tree}, after the formal mechanics of the model are introduced.}

Here, only a single round is needed to achieve full revelation and efficiency. In general, however, the firm may only slowly reveal, using the successive proposals of the regulator to steer the final choice towards its preferred outcome. To see this, consider the same example except that ex-ante the regulator is now also unaware of $\xx$ (i.e., only aware of $\yy$); $\xx$ and $\{\aa,\bb\}$ can be revealed independently. 

Consider what happens when only a single round of revelation is permitted: initially the regulator shortlists the merger $\yy$, as it is the only merger it can express. If the firm reveals everything, the regulator will choose $\bb$, which is preferred (by both players) to the outside option, $\yy$. This is efficient, and indeed a Pareto improvement over no revelation, but is not the best the firm can do. Instead of fully revealing, the firm can reveal $\xx$ alone, which would be chosen and which the firm prefers to $\bb$. 

Notice that after this first round of revelation, we find ourselves in the state-of-affairs described at the \emph{beginning} of the first example: the regulator is aware of $\xx$ and $\yy$ and has shortlisted $\xx$. And so, another round of interaction will lead to the same efficient and fully revealing outcome.

\begin{figure}
\begin{minipage}{0.24\textwidth}
\centering
\begin{tikzpicture}[scale=.5]
  \foreach \x in {0,1,2,3,4} \draw[gray, thin, dotted] (\x,-1) -- (\x,5);
  \foreach \y in {0,1,2,3,4} \draw[gray, thin, dotted] (-1,\y) -- (5,\y);
  \draw[dotted,thick,gray!30] (4.12,0) arc (-.2:26:4.5);
    \draw[dotted,thick] (4.12,0) arc (-.2:86:4.5);
    
    \draw[thick,->] (-1,0) -- (5,0) node[anchor=north] {$z_c$};
    \draw[thick,->] (0,-1) -- (0,5) node[anchor=east] {$z_f$};

    
    \fill[lam5](0,0) circle (5pt) node[anchor=south west] {\yy};

    \fill[gray!30] (2,2) circle (5pt) node[anchor=south west] {\texttt{x}};
    \fill[gray!30] (3,3) circle (5pt) node[anchor=south west] {\texttt{a}};
    \fill[gray!30](4,1) circle (5pt) node[anchor=south west] {\texttt{b}};

    \node[star, star points=5, star point ratio=2.25, fill=lamy, minimum size=12pt, inner sep=0pt] at (0,0) {};

    \node[color=black] at (2,-2) {$f(\{\yy\}) = \yy$};
\end{tikzpicture}
\end{minipage}
\hfill
\begin{minipage}{0.24\textwidth}
\centering
\begin{tikzpicture}[scale=.5]
  \foreach \x in {0,1,2,3,4} \draw[gray, thin, dotted] (\x,-1) -- (\x,5);
  \foreach \y in {0,1,2,3,4} \draw[gray, thin, dotted] (-1,\y) -- (5,\y);
  \draw[dotted,thick,gray!30] (4.12,0) arc (-.2:26:4.5);
    \draw[dotted,thick] (4.12,0) arc (-.2:86:4.5);
    
    \draw[thick,->] (-1,0) -- (5,0) node[anchor=north] {$z_c$};
    \draw[thick,->] (0,-1) -- (0,5) node[anchor=east] {$z_f$};
    \fill[lamy](0,0) circle (5pt) node[anchor=south west] {\yy};
    \fill[lamx] (2,2) circle (5pt) node[anchor=south west] {\xx};

    \fill[gray!30] (3,3) circle (5pt) node[anchor=south west] {\texttt{a}};
    \fill[gray!30](4,1) circle (5pt) node[anchor=south west] {\texttt{b}};

    \node[star, star points=5, star point ratio=2.25, fill=lamx, minimum size=12pt, inner sep=0pt] at (2,2) {};

    \node[color=black] at (2,-2) {$f(\{\yy, \xx\}) = \xx$};
\end{tikzpicture}
\end{minipage}
\hfill
\begin{minipage}{0.24\textwidth}
\centering
\begin{tikzpicture}[scale=.5]
  \foreach \x in {0,1,2,3,4} \draw[gray, thin, dotted] (\x,-1) -- (\x,5);
  \foreach \y in {0,1,2,3,4} \draw[gray, thin, dotted] (-1,\y) -- (5,\y);
  \draw[dotted,thick,gray!30] (4.12,0) arc (-.2:26:4.5);
    \draw[dotted,thick] (4.12,0) arc (-.2:86:4.5);
    
    \draw[thick,->] (-1,0) -- (5,0) node[anchor=north] {$z_c$};
    \draw[thick,->] (0,-1) -- (0,5) node[anchor=east] {$z_f$};
    \fill[lamy](0,0) circle (5pt) node[anchor=south west] {\yy};
     \fill[lama] (3,3) circle (5pt) node[anchor=south west] {\aa};
     \fill[lama](4,1) circle (5pt) node[anchor=south west] {\bb};

    \fill[gray!30] (2,2) circle (5pt) node[anchor=south west] {\texttt{x}};

    \node[star, star points=5, star point ratio=2.25, fill=lama, minimum size=12pt, inner sep=0pt] at (4,1) {};

    \node[anchor=center, color=black] at (2,-2) {$f(\{\yy, \aa, \bb\}) = \bb$};
\end{tikzpicture}
\end{minipage}
\hfill
\begin{minipage}{0.24\textwidth}
\centering
\begin{tikzpicture}[scale=.5]
  \foreach \x in {0,1,2,3,4} \draw[gray, thin, dotted] (\x,-1) -- (\x,5);
  \foreach \y in {0,1,2,3,4} \draw[gray, thin, dotted] (-1,\y) -- (5,\y);
  \draw[dotted,thick,gray!30] (4.12,0) arc (-.2:26:4.5);
    \draw[dotted,thick] (4.12,0) arc (-.2:86:4.5);
    
    \draw[thick,->] (-1,0) -- (5,0) node[anchor=north] {$z_c$};
    \draw[thick,->] (0,-1) -- (0,5) node[anchor=east] {$z_f$};
    \fill[lamy](0,0) circle (5pt) node[anchor=south west] {\yy};
    \fill[lamx] (2,2) circle (5pt) node[anchor=south west] {\xx};
     \fill[lama] (3,3) circle (5pt) node[anchor=south west] {\aa};
     \fill[lama](4,1) circle (5pt) node[anchor=south west] {\bb};


    \node[star, star points=5, star point ratio=2.25, fill=lama, minimum size=12pt, inner sep=0pt] at (3,3) {};

    \node[anchor=center, color=black] at (2,-2) {$f(\{\yy, \xx, \aa, \bb\}) = \aa$};
\end{tikzpicture}
\end{minipage}
\caption{The regulators most preferred efficient and incentive compatible profile of outcomes.}
\label{fig:selection_function}
\end{figure}

The regulator, by playing a greedy strategy, i.e., always shortlisting its preferred merger across those that the firm would accept, induces the same outcomes as the following direct mechanism:
$$
f(\{\yy\}) = \yy, \qquad f(\{\yy,\xx\}) = \xx, \qquad f(\{\yy,\aa,\bb\}) = \bb,\qquad f(\{\yy, \xx, \aa,\bb\}) = \aa
$$
as shown in Figure \ref{fig:selection_function}. It is straightforward to see that this point-wise dominates any other efficient and incentive compatible profile of outcomes. 

Theorem \ref{thm:recvbestMech} shows that greedy optimization in the IRP is at least as good for the regulator as any other efficient and incentive compatible mechanism: the IRP constitutes thus constitutes the optimal regulatory framework.  Thus, rather than committing to a particular welfare \emph{standard}, the optimality of which will depend critically on the set of mergers the firm can reveal, the regulator is better off committing to a particular negotiation \emph{process}.
Theorem \ref{thm:recvbestMech} reinforces the conclusions of \cite{nocke2010dynamic}, who show that greedy strategies will maximize the regulator's discounted expected utility within the specific context of Cournot competition where merger opportunities arise stochastically over time.

\section{Model}
\label{sec:model}

\setlength{\abovedisplayskip}{10pt}
\setlength{\belowdisplayskip}{10pt}

The environment is described by a tuple $(\A, (u_d,u_e), \Rev)$ where

\begin{itemize}
  \item $\A$ is a set of projects, the universe of actions that can be taken by the decision maker,
  \item $u_i : \A \to \R$ are the players'  utility functions,\footnote{
  \label{ft:ordinal}
  None of the results in this paper rely on the cardinal information contained in the utility functions. 
The model is specified with utility functions, rather than ordinal preference relations, only to keep the examples and notation simple and close to the literature on project choice / mechanism design. That everything that follows could be rewritten in terms of ordinal preference data reinforces the assumption that utility is not transferable and utility values are uncontactable.
  }
  \item $\Rev$ is a collection of non-empty subsets of $\A$, referred to as \emph{revelation types}, with a minimal (with respect to set inclusion) element, $r^\dag \in \Rev$.
\end{itemize}
In the event that $\A$ is infinite, I further require that (i) $\A$ entertains some topology, (ii) $u_i$ is continuous and bounded, and (iii) each $r \in \Rev$ is a compact subset of $\A$ such that the set $\{r' \in \Rev \mid r' \subseteq r\}$ is finite.

The novelty of the model arises from $\Rev$, which captures the set of possible revelations a player can make as described in detail below. 
In particular, a player of type $r \in \Rev$ cannot express any project $a \notin r$ (i.e., cannot communicate or reveal $a$ to other agents, describe or commit to $a$ in a contract, etc.).  Inexpressibility could arise for a variety of interrelated reasons---unawareness of events or outcomes, the inability to articulate them, legal restrictions of what can be enforced, technological constraints, etc.\footnote{The model is agnostic to the exact interpretation of expressibility: It may be possible for a player to internally reason about those things she cannot express, for example, a decision maker might think ``If the expert reveals some project with such and such features, I will enact it.'' Of course, the \emph{actual} project in question cannot be expressed, so even though the decision maker might be able to abstractly envision it, it is not (yet) a feasible choice. A more severe form of inexpressibility, e.g., total unawareness, might preclude such internal ruminations.}
For $r,r' \in \Rev$, say that $r$ is \emph{more expressive} than $r'$ (or that $r'$ is \emph{less expressive} than $r$), if $r' \subseteq r$. 


Within this setup, I maintain three assumptions about the role of expressibility:

\begin{assumption}[Expertise]
\label{ass:exp}
The decision maker is the minimally expressive type $r^\dag$. 
\end{assumption}

\noindent Since the expert can potentially be of any type in $\Rev$, Assumption \ref{ass:exp} is tantamount to the assumption that it is commonly known that the expert is more expressive than the decision maker. 

Second, to capture information spillover, I assume that the revelation types of a players determine not only the full set of projects they can express, but also subsets of projects that can be independently revealed. 

\begin{assumption}[Information Spillover]
\label{ass:spill}
The set possible revelations that can be made by a type $r$ is exactly $\{r' \in \Rev \mid r' \subseteq r\}$.
\end{assumption}

To see how this represents information spillover, consider two projects $a$ and $b$ that use the same (hitherto unknown to decision maker) technology, revealing either project might inadvertently reveal the other. This would be captured by a state-space $\Rev$ such that for all $r \in \Rev$, $a \in r$ if and only if $b \in r$; then, under Assumption \ref{ass:spill}, $a$ and $b$ can only ever be revealed jointly. The structure of $\Rev$ can reflect different epistemic or technological constraints relevant to the environment. The condition that $\{r' \in \Rev \mid r' \subseteq r\}$ is finite for all $r \in \Rev$ requires that for each type of expert there are a finite number of revelations a player of type $r$ could make.

Finally,  I assume that revelation is voluntary in the sense that withholding the existence of projects is undetectable, and so, cannot be punished.

\begin{assumption}[Voluntary Revelation]
\label{ass:vol}
A revelation type $r$ can act exactly as any $r' \in \Rev$ such that $r' \subseteq r$ without detection.
\end{assumption}

For some results, it will be useful to consider \emph{rich} environments, where there are limited common knowledge restrictions on what has yet to be revealed. In a rich environment, the decision maker can never rule out the existence of further projects yielding (essentially) arbitrary payoff profiles. Formally, call the environment \emph{rich} if for any type $r \in \Rev$, and non-empty open intervals  $I^{d}, I^{e} \subseteq \R$ there exists a project $a^\circ \in \A$ such that $r \cup \{a^\circ\} \in \Rev$ and $u_i(a^\circ) \in I^{i}$ for $i \in \{d,e\}$. While it is more cumbersome, it is sufficient to capture richness in an ordinal manner by taking the endpoints of the intervals to be delineated by projects in $r$.

\subsection{Selection Rules}
\label{sec:selection_rules}
 \def\:{\!:\!}

If the decision maker were unencumbered by expressibility or commitment issues, he could implement outcomes via direct mechanisms. In the environment, a direct mechanism is a \emph{selection rule}, a mapping from revelation types to elements thereof:

\begin{equation*}
f: \Rev \to \A
\end{equation*}
where $f(r) \in r$ for all $r \in  \Rev$. Let $\F$ denote the set of all selection rules.

Given Assumption \ref{ass:vol}---that a type has an inalienable ability to reveal \emph{less}---incentive compatibility of a selection rule is tantamount to requiring that the expert does not strictly prefer the outcome selected from some less expressive type. 

\begin{definition}
\label{def:fullyreavealing}
Call a selection rule $f \in \F$ \df{monotone} if for $r, r' \in \Rev$
\begin{equation}
\label{eq:fll_reavealing}
\tag{\textsc{m}}
 r' \subseteq r\quad \text{ implies } \quad u_e(f(r')) \leq u_e(f(r))  
\end{equation}
and \df{strictly monotone} if \eqref{eq:fll_reavealing} holds strictly whenever $f(r') \neq f(r)$. Let $\F^{m}\subseteq \F$ denote the set of monotone selection rules and $\F^{sm} \subseteq \F^{m}$ the strictly monotone selection rules.
\end{definition}

While it is clear that a selection rule that fails to be monotone could never be induced by any strategic interaction that relies on voluntary disclosure, the converse is not  immediately obvious. Even under monotonicity, it is unclear how a decision maker (or any other mechanism designer) can induce $f$ given his limited expressibility. The decision maker cannot simply announce what will be offered contingent on each revelation, as by design, he cannot articulate anything more expressible than his own type. That is to say, the decision maker cannot in general describe $f(r)$ until $r$ has been revealed. This is not a problem of incentives but rather a mechanical constraint governed by the nature of expressibility.

As we will see in the next section, monotonicity turns out to be sufficient to guarantee that a selection rule can be effected via strategic means: in particular, every monotone selection rule can be implemented by some strategy in the iterated revelation protocol.

  \subsection{The Iterated Revelation Protocol}
  \label{sec:IRP}

Using the language of revelation types, we can now formalize the procedure from Section \ref{sec:ex}, in which the decision maker and expert take turns shortlisting and revealing sets of projects, respectively. 
This repeated shortlist/revelation interaction is captured by the \emph{iterated revelation protocol} (IRP):

\begin{enumerate}[leftmargin=2cm]
   \item[\textsc{initial step}\ ---]\label{irm1}  The initial revelation is $r_0 = r^\dag$ and decision maker shortlists a project $p_0 \in r_0$.
   \item[\textsc{iterative step}\ ---]\label{irm2}  Given $(r_0, p_0, \ldots, r_n, p_n)$, the sequence of alternating revelations/shortlists (ending in a shortlist proposal): The expert then reveals $r_{n+1} \in \Rev$. 
    \begin{itemize}
      \item If $r_{n} \subsetneq r_{n+1}$, then the decision maker adds $p_{n+1} \in r_{n+1}$ to the shortlist and the \textsc{iterative step} is repeated.
      \item Otherwise, the protocol moves to the \textsc{final step}
    \end{itemize}
   \item[\textsc{final step}\ ---]\label{irm3} The expert chooses project $a \in \{p_0, \ldots, p_n\}$.
 \end{enumerate}

A critical feature of the IRP is that the decision maker only ever shortlists contracts that he can express at the time of the proposal. Thus, the IRP is straightforward to carry out---the procedure and commitment structure can be described ex-ante, and each proposal in the sequence can be expressed given what has already been revealed up to that point. 

 \subsection{The IRP as a Dynamic Game}
 \label{sec:game}

In this section, I show how the dynamic environment dictated by the IRP---wherein the players take turns revealing / shortlisting projects---can be represented formally as a dynamic game.\footnote{Although this formalization clarifies the strategic details of the IRP (thus simplifying the discussion of game-theoretic constructions, e.g., best responses, etc) it also invites the potentially unwanted side effect of encouraging all further analysis to be through the lens of equilibrium. As such, I caution that while it is possible to formalize the IRP as a game form, this does not mean that the players (in particular the decision maker) know what this game form looks like; in particular, he will be unable to express/conceive of all histories in his current information set and therefore unable to engage in backwards induction reasoning or otherwise entertain well defined beliefs over future actions. Similar issues arise in the (small) literature on games with unawareness \cite{feinberg2020games,heifetz2013dynamic}.} Towards making this formal, define a \emph{public history} as a finite sequence of alternating revelations and additions to the shortlist: $(r_0, p_0, r_1, p_1,\ldots)$ where $r_0 = r^\dag$ and for each $m$,  $r_m \in \Rev$ and $p_m \in r_m$. Moreover, if $m > 0$, then $r_{m-1} \subsetneq r_{m}$; that is each successive revelation is \emph{strictly} more expressive. A public history can end in either a shortlist addition, $p_i$, or a revelation of new projects, $r_i$. Let $\H$ denote the set of public histories and $\H^{P}$ and $H^{R}$ those ending in a shortlist proposal and a revelation, respectively.

\begin{figure}[]
\centering
\begin{tikzpicture}[
  level distance=1cm,
  every node/.style={
  draw,
  rectangle,
  rounded corners,
  minimum width=10mm,
  minimum height=6mm,
  inner sep=2pt,
  font=\scriptsize
},
  edge from parent/.style={draw, -latex}
  ]

\node[draw=lam4, line width=1.6pt,] {$\emptyset$}
  child[thick, draw=black, sibling distance=5.5cm] {node[name=r1] {$r_e {=} \{\xx,\yy\}$}
    child[thick, draw=black, sibling distance=1.5cm] {node {$p_0 {=} \xx$}
      child {node[circle, draw, minimum size=6mm, inner sep=0pt] {$\xx$}}}
    child[thick, draw=black, sibling distance=1.5cm] {node {$p_0 {=} \yy$}
      child {node[circle, draw, minimum size=6mm, inner sep=0pt] {$\yy$}}}
  }
  child[draw=lam4, line width=1.6pt, sibling distance=5.5cm] {node[name=r2] {$r_e {=} \{\xx,\yy,\aa,\bb\}$}
    child[draw=lam4, line width=1.6pt,  sibling distance=6.5cm] {node {$p_0 {=} \xx$}
      child[thick, draw=black, sibling distance=0cm] {node[circle, draw, minimum size=6mm, inner sep=0pt] {$\xx$}}
      child[draw=lam4, line width=1.6pt, sibling distance=4cm] {node {$r_1 {=} \{\xx,\yy,\aa,\bb\}$}
        child[thick, draw=black, sibling distance=3cm, level distance=1.5cm] {node {$p_1 {=} \xx,$}
        child[thick, draw=black, sibling distance=0cm, level distance=1cm] {node[circle, draw, minimum size=6mm, inner sep=0pt] {$\xx$}}}
        child[thick, draw=black, sibling distance=6cm, level distance=1.5cm] {node {$p_1 {=} \yy,$}
        child[thick, draw=black, sibling distance=.75cm, level distance=1cm] {node[circle, draw, minimum size=6mm, inner sep=0pt] {$\xx$}}
            child[thick, draw=black, sibling distance=.75cm, level distance=1cm] {node[circle, draw, minimum size=6mm, inner sep=0pt] {$\yy$}}}
        child[draw=lam4, line width=1.6pt,  sibling distance=-3cm, level distance=1.5cm] {node {$p_1 {=} \aa,$}
        child[thick, draw=black, sibling distance=.75cm, level distance=1cm] {node[circle, draw, minimum size=6mm, inner sep=0pt] {$\xx$}}
            child[draw=lam4, line width=1.6pt, sibling distance=.75cm, level distance=1cm] {node[circle, draw, minimum size=6mm, inner sep=0pt] {$\aa$}}}
        child[thick, draw=black, sibling distance=0cm, level distance=1.5cm] {node {$p_1 {=} \bb,$}
        child[thick, draw=black, sibling distance=.75cm, level distance=1cm] {node[circle, draw, minimum size=6mm, inner sep=0pt] {$\xx$}}
            child[thick, draw=black, sibling distance=.75cm, level distance=1cm] {node[circle, draw, minimum size=6mm, inner sep=0pt] {$\bb$}}}
      }
    }
    child[thick, draw=black, sibling distance=6.5cm] {node {$p_0 {=} \yy$}
      child[thick, draw=black, sibling distance=4.5cm] {node {$r_1 {=} \{\xx,\yy,\aa,\bb\}$}
            child[thick, draw=black, sibling distance=0cm, level distance=1.5cm] {node {$p_1 {=} \xx,$}
            child[thick, draw=black, sibling distance=.75cm, level distance=1cm] {node[circle, draw, minimum size=6mm, inner sep=0pt] {$\yy$}}
            child[thick, draw=black, sibling distance=.75cm, level distance=1cm] {node[circle, draw, minimum size=6mm, inner sep=0pt] {$\xx$}}
            }
            child[thick, draw=black, sibling distance=-3cm, level distance=1.5cm] {node {$p_1 {=} \yy,$}
            child[thick, draw=black, sibling distance=0cm, level distance=1cm] {node[circle, draw, minimum size=6mm, inner sep=0pt] {$\yy$}}
            }
            child[thick, draw=black, sibling distance=6cm, level distance=1.5cm] {node {$p_1 {=} \aa,$}
              child[thick, draw=black, sibling distance=.75cm, level distance=1cm] {node[circle, draw, minimum size=6mm, inner sep=0pt] {$\yy$}}
              child[thick, draw=black, sibling distance=.75cm, level distance=1cm] {node[circle, draw, minimum size=6mm, inner sep=0pt] {$\aa$}}}
           child[thick, draw=black, sibling distance=3cm, level distance=1.5cm] {node {$p_1 {=} \bb,$}
             child[thick, draw=black, sibling distance=.75cm, level distance=1cm] {node[circle, draw, minimum size=6mm, inner sep=0pt] {$\yy$}}
            child[thick, draw=black, sibling distance=.75cm, level distance=1cm] {node[circle, draw, minimum size=6mm, inner sep=0pt] {$\bb$}}}
        }
      child[thick, draw=black, sibling distance=0cm] {node[circle, draw, minimum size=6mm, inner sep=0pt] {$\yy$}}
    }
  };

\node[font=\scriptsize, draw=none] at (-4.5,0) {N};
\node[font=\scriptsize, draw=none] at (-4.5,-1) {DM};
\node[font=\scriptsize, draw=none] at (-4.5,-2) {E};
\node[font=\scriptsize, draw=none] at (-4.5,-3) {DM};
\node[font=\scriptsize, draw=none] at (-4.5,-4.5) {E};

\draw[dotted, thick] (r1) -- (r2);

\end{tikzpicture}
\caption{A game tree depicting the example from Section \ref{sec:ex} with type space $\Rev = \{\{\xx, \yy\}, \{\xx, \yy, \aa, \bb\}\}$. The active player at each row is specified to the left, and terminal histories are demarcated by circular nodes. The interaction analyzed in the example follows the highlighted sequence of play.}
\label{fig:game-tree}
\end{figure}

 At the outset of the game, nature chooses the type of expert $r_e \in \Rev$. Each subsequent (non-terminal) history of the game is defined by a nature's incipient choice and a public history. The initial player-active history is $(r_e, r_0)$, where $r_0 = r^\dag$; the decision maker is active, and has an action set $r_0$. That is, he can choose to shortlist a project $p_0 \in r_0$, moving to history $(r_e, r_0,p_0)$. This corresponds to the decision maker's initial proposal in the \textsc{initial step} of the IRP.

At any history ending in a revelation, $(r_e, r_0, p_0, \ldots, r_n) \in \H^{R}$, the decision maker is active and has an action set given by $r_n$, that is, he can choose some $p_n \in r_n$. The new history is then $(r_e, r_0, p_0, \ldots, r_n, p_n)$. This corresponds decision maker's addition to the shortlist in the \textsc{iterative step}.

At any history ending in an addition to the shortlist, $(r_e, r_0, p_0, \ldots, r_n, p_n) \in \H^{P}$ the expert is active. She can take two distinct types of actions: she can either make a revelation, $r_{n+1} \in \Rev$ such that $r_n \subsetneq r_{n+1} \subseteq r$ moving to $(r_e, r_0, p_0, \ldots, r_n, p_n, r_{n+1})$, or end the game by choosing an project $a \in \{p_0 ,\ldots, p_n\}$, which moves to the terminal history $(r_e, r_0, p_0, \ldots, r_n, p_n, a)$, where payoffs are $u_d(a)$ and $u_e(a)$.
These different types of actions correspond to the expert's revelation in the \textsc{iterative step} and choice of project in the \textsc{final step}, respectively. 

Information sets are given by the players' information about nature's choice. Specifically, the sequence of revelations/shortlist additions are observed by both players, but the decision maker can not distinguish between histories that differ only in nature's choice of the expert's revelation type. Figure \ref{fig:game-tree} depicting the dynamic game formalizing the initial example in Section \ref{sec:ex}, where the type space is $\Rev = \{\{\xx, \yy\}, \{\xx, \yy, \aa, \bb\}\}$. Notice, at the decision maker's information set in the second layer (before anything has been revealed by the expert), the decision maker may not be able to envision the continuation game tree nor entertain well defined beliefs about continuation play.  

 \begin{definition}
\label{def:strat-dm}
Let $s$ denote a \textbf{strategy for the decision maker} and $\sigma$ denote a \textbf{strategy for the expert of type $\bm{r}$}. These
 determine the players' actions at each history at which they are active, parameterized by the set of public histories. In particular:
\begin{itemize}
  \item  $s: \H^{R} \to \A$: for each relevant public history, $h = (r_0, p_0, \ldots, r_n) \in \H^{R}$, we have $s(h) \in r_n$; the decision maker's strategy selects which project is added to the shortlist
  \item $\sigma: \H^{P} \to \Rev \cup \A$: for each relevant public history, $h = (r_0, \ldots, r_n, p_n) \in \H^{P}$, we have
  \begin{itemize}
        \item  $\sigma(h) \in \{ r' \in \Rev \mid r_n \subsetneq  r' \subseteq r\}$; the expert's strategy selects her revelation should she continue revealing, or,
        \item  $\sigma(h) \in  \{p_0, \ldots, p_n\}$; the expert's strategy selects the enacted project, should she stop revealing and end the game.
    \end{itemize}
 \end{itemize}
Let $S$ and $\Sigma(r)$ denote the set of strategies available to the decision maker and an expert of type $r$, respectively.
\end{definition}



A few notes on strategies.
First, each pair of strategies $(s,\ss) \in S \times \Sigma(r)$  jointly determine a unique project that will be implemented.\footnote{While I restrict attention to pure strategies, this does \emph{not} preclude the use of random allocation rules, which can be incorporated explicitly by requiring that $\A$ is a convex space and each $r \in \Rev$ is a convex subset of $\A$.} Denote this project by $a(s,\ss) \in \A$. Second, while strategies determine play at every history, only histories that are expressible by the players will ever be reached. In particular, notice that for a fixed type of expert $r$, the available revelations $\{r' \in \Rev \mid r_n \subsetneq r' \subseteq r \}$ will be empty after a finite number of rounds; after this, the expert must end the game.  Finally, the results in this paper do not require strategies that condition on the full history. Indeed, the main strategic constructions of this paper (greedy and locally rational strategies) find the decision maker conditioning his choice only on the \emph{most recent} addition to the shortlist.  

 \begin{definition}
\label{def:br}
Call $\ss \in \Sigma(r) $ a \textbf{best response (for revelation type $\bm{r}$)} to $s \in S$ if 
 \begin{equation}
\label{eq:BR}
 u_e(a(s,\ss)) \geq u_e(a(s,\ss'))  \qquad \text{ for all } \quad \sigma ' \in \Sigma(r) .
  \tag{\textsc{br}}  
\end{equation}
Let $\BR(s, r) \subseteq \Sigma(r)$ denote the set of all best responses to $s$,
and $\BRR(s, r)$ those that are fully revealing (i.e., which reach a history where $r$ is revealed).
\end{definition}

\subsection{Implementation}

I now show that the IRP can serve as a universal mechanism, in the sense that any monotone selection rule can be implemented via the appropriate choice of strategy. Formally:

 \begin{definition}
\label{def:implements}
Say that the strategy $s \in S$ \df{implements} (resp., \df{transparently implements}) the selection rule $f \in \F$ if for all $r \in \Rev$
$$
f(r) = a(s,\ss) \quad  \text{for some} \quad  \ss \in \BR(s,r) \quad (\text{resp., $\ss \in \BRR(s,r)$})
$$
\end{definition}  

The following theorem shows that a sufficiently sophisticated decision maker need not be constrained by his expressibility. That is, despite the fact that the IRP never requires committing to a project before it has been revealed, the protocol can implement any monotone selection rule. As such, any direct mechanism can be effectively captured via the appropriate choice of strategy in the IRP.


\begin{theorem}
\label{thm:mondisclosure}
Let $f \in \F$. The following are equivalent: 
\begin{enumerate}
  \item \label{md:ic} $f$ is monotone,
  \item \label{md:im} $f$ is implemented by some $s \in S$,
  \item \label{md:*} $f$ is  transparently implemented by some $s \in S$.
\end{enumerate}
\end{theorem}

Of course, that a strategy $s$ implements the selection rule $f$, does not guarantee the outcome will be determined by $f$, since there may be many best responses for the expert that result in different enacted projects. A stronger condition would ask that $f$ is the unique outcome of the IRP. 

\begin{definition}
\label{def:full_implements}
Say that the strategy $s \in S$ \df{fully implements}  the selection rule $f \in \F$ if for all $r \in \Rev$
$$
f(r) = a(s,\ss) \quad  \text{for all} \quad  \ss \in \BR(s,r) 
$$
\end{definition}  

\noindent The theorem below shows that full implementation is possible exactly when $f$ is \emph{strictly monotone}. Full implementation is important because it eliminates any ambiguity about the outcome of the interaction, conditional on the expert's type.

 \begin{theorem}
\label{thm:full_implements}
A selection rule $f \in \F$ is strictly monotone if and only if there exists a strategy $s \in S$ that fully implements it.
\end{theorem}

\section{Efficient Outcomes}
\label{sec:eff}

Theorem \ref{thm:mondisclosure} ensures the existence of some strategy, $s$, to implement any arbitrary selection rule, $f$. However, envisaging such a strategy may require a rather complete picture of the payoff consequences of potential revelations. Further still, calculating the \emph{optimal} selection rule would require formulating precise beliefs about the likelihood of these potential revelations.  These demands are burdensome for decision makers who suffer under extreme uncertainty, entertaining an incomplete and shrouded concept of what might yet be revealed.

This section shows that, when the decision maker must implement an efficient outcome, these limitations impose no practical constraints: Greedy optimization, a simple heuristic strategy that does not require any forward-looking behavior, implements the decision maker's ex-post optimal efficient selection rule. As such, under the dictate of efficiency, there is effectively zero cost to epistemic obstructions such as unawareness.

\begin{definition}
\label{def:mechPO}
Call a selection rule $f \in \F$ \df{efficient} if for all $r \in \Rev$, there is no $a \in r$ such that 
$$u_d(a) \geq u_d(f(r)) \qquad \text{ and } \qquad u_e(a) \geq u_e(f(r)),$$ 
with at least one inequality holding strictly.
\end{definition}  

Optimality is required ex-post, so an efficient selection rule results in Pareto optimal use of the revealed projects. 

\subsection{Characterizing Efficiency}

To formalize greedy optimization, I first introduce the idea of an incentive compatible addition to the shortlist. At a given public history, the set of incentive compatible additions is the set of projects that could be chosen over all projects already on the shortlist.  

\begin{definition}
\label{def:ICset}
Given a public history $h = (r_0, p_0, \ldots, r_n) \in \H^{R}$, define
\begin{equation}
\label{eq:ICset}
\IC(h) = \{a \in r_n \mid u_e(a) \geq u_e(p_i), \text{ for all } i < n \}.
\end{equation}
as the set of projects that could be chosen if added to the shortlist.
\end{definition}  

\noindent With this definition in place, I can now define what it means for a decision maker's strategy to be (mostly) greedy:

\begin{definition}
\label{def:mostly_greedy}
Call $s \in S$ \df{mostly greedy} if at every public history $h = \H^{R}$,
\begin{enumerate}[label=\textsc{\textup{(mg\arabic*)}}]
  \item\label{mgr:1}  $s(h) \in \IC(h)$, 
  \item\label{mgr:2} for all $a \in \IC(h)$, $u_d(s(h)) \geq u_d(a)$, and,
  \item\label{mgr:3} if $a \in \IC(h)$ and $u_d(s(h)) = u_d(a)$ then $u_e(s(h)) \geq u_e(a)$.
\end{enumerate}
\end{definition}  

A mostly greedy strategy has three dictates that must hold at every history $h$: first, \scref{mgr:1} requires that the decision maker's addition to the shortlist $s(h)$ is an incentive compatible addition so that it might be chosen if the IRP ended with no further additions; second \scref{mgr:2} requires that the decision maker greedily optimizes so that $s(h)$ maximizes his payoff (over all other incentive compatible additions); finally, \scref{mgr:3} requires that the decision maker break ties in favor of the expert's payoff. This last mandate is why such strategies are only \emph{mostly} greedy, and what ensures that the outcome will be efficient.

\begin{theorem}
\label{thm:recvbestMech}
Let $s \in S$ be mostly greedy. Then $s$ implements the decision maker's preferred efficient selection rule $f^\star \in \F$. That is, if $f'$ is any other monotone and efficient selection rule, then for all  $r \in \Rev$, $u_d(f^\star(r)) \geq u_d(f'(r))$.
\end{theorem}

Theorem \ref{thm:recvbestMech} shows that the among all selection rules that are both implementable (i.e., monotone) and efficient, the ex-post decision-maker-optimal project is enacted via mostly greedy optimization. This strategy is both simple to describe and to compute and requires no forward planning whatsoever.\footnote{The `dual' strategy, which instead delegates all decision making power to the expert, delineates the opposite bounds selecting the expert's most preferred selection rule. This---rather boring---strategy simply selects out of each revelation the expert's most preferred project subject to an individual rationality constraint.}

Notice also that because greedy optimization ex-post dominates any other efficient mechanism, there is never any incentive for the decision maker to deviate from greedy optimization. As such, the only commitment power needed is the commitment to the IRP itself. Within the context of antitrust regulation, this indicates that the regulator need only construct the appropriate regulatory framework to facilitate the IRP; given this commitment, there will be no strategic uncertainty about how the regulator will behave.

\subsection{Comparative Statics}

Given the appeal of mostly greedy optimization---both in terms of simplicity and also in the outcomes effected---I will now examine how the resulting selection rule depends on the initial expressibility of the players.  
Theorem \ref{thm:compStatics} has four parts: The first two claims relate the players initial expressibility given a fixed revelation type space; the second two claims relate the expert's `articulation,' that is, her ability to stymie information spillover.

\begin{theorem}
\label{thm:compStatics}
Let $\Rev$ and $\Qev$ denote two revelation types spaces defined on the same underlying set of projects. Let $f^{\Rev}$ and $f^{\Qev}$ denote the decision maker's preferred efficient, monotone selection rules (i.e., those implemented by the respective mostly greedy strategies). Then
\begin{enumerate}
  \item\label{thm:cs:better_for_d}  If $\Qev = \{r' \in \Rev \mid r \subseteq r'\}$ for some $r \in \Rev$, then $u_d(f^{\Rev}(q)) \leq u_d(f^{\Qev}(q))$ for all $q \in \Rev \cap \Qev$.
   \item Let $r, r' \in \Rev$ with $r' \subseteq r$, the sign of the difference between $u_d(f^{\Rev}(r))$ and $u_d(f^{\Rev}(r'))$ is not determined.
   \item Let $\Qev \subseteq \Rev$ be such that  $r^\dag = q^\dag$. Further let $r \in \Rev \cap \Qev$. Then $u_d(f^{\Rev}(r)) \leq u_d(f^{\Qev}(r))$.
   \item If $\Rev = 2^\A$, then (up to indifference) there is a unique selection rule that is efficient, monotone and individually rational.\footnote{Individual rationality requires  that $u_d(f(r)) \geq \max_{a \in r^\dag} u_d(a)$ for all $r\in \Rev$.}
\end{enumerate}
\end{theorem}

Claim (1) states that a decision maker who is initially of a more expressive type will do better. Although unsurprising, it clarifies that the IRP (under mostly greedy optimization) will work even when the exact expressivity of the decision maker is not commonly known (that is, if the decision maker's and expert's types, $r_d$ and $r_e$ were arbitrary elements of $\Rev$ such that $r_d \subseteq r_e$). In such a case, the decision maker could first announce his type, and then run the IRP given the remaining types: $\Qev = \{r' \in \Rev \mid r_d \subseteq r'\}$; claim (1) guarantees the decision maker finds it in his best interest to announce his true type, as his eventual payoff is monotone in his announcement. 

Claim (2), also unsurprisingly, notes that the decision maker can be both hurt or helped by the novel projects expert can express. This is clear from the example in Section \ref{sec:ex}: for the selection rule $f$ resulting from mostly greedy optimization, shown in Figure \ref{fig:selection_function}, we have $u_d(f(\{\yy\})) < u_d(f(\{\yy, \aa,\bb\}))$ and also $u_d(f(\{\yy, \aa,\bb\})) > u_d(f(\{\yy, \xx, \aa,\bb\}))$.

Claims (3) and (4) relate outcomes to the expert's `articulation' as measured by the fineness of $\Rev$. As the revelation type space becomes finer, i.e., includes more subsets of $\A$, an expert of a given expressivity has more possible revelations. This can be understood as being better able to prevent information spillover, since the more articulate expert can reveal more distinct subsets of projects. Claim (3) shows that the decision maker always does (weakly) better when the expert is more inarticulate (i.e., when the type space is coarser). Claim (4) examines the limit, where the expert can reveal any project independently, and shows that the expert gets her most preferred project enacted (subject to an individual rationality constraint ensuring the decision maker is no worse off than what he could achieve alone).

Taken together, claims (3) and (4) resolve a complete picture of how information spillover drives the division of surplus. When the expert is fully articulate, she receives the full value of the surplus garnered by her novel projects. As she becomes less and less articulate, unable to precisely reveal only the projects she wants to, the expert cedes rents back to the decision maker. At the other limit, when the expert can reveal only everything or nothing, the decision maker receives the full surplus, as he is then able to get his most preferred project enacted
(again subject to an individual rationality constraint ensuring the expert is willing to reveal).

\section{General Strategic Analysis}
\label{sec:general_strategic_analysis}

There are many strategies in the IRP; when constrained to choose efficient outcomes, the decision maker's choice of strategy is straightforward, but what about when there is no requirement of efficiency? 
In general, when adding a project to the shortlist, the decision maker faces a trade-off between (i) maximizing his own payoff should that project get chosen by the expert, and (ii) minimizing the expert's payoff so as to relax future incentive constraints as much as possible.

For example, for the first $n-1$ revelations, the decision maker could add the expert's least preferred project, and subsequently, shortlist the project that maximizes his own payoff. Should the interaction persist to the $n^{\text{th}}$ round, such a strategy might reward the decision maker handsomely as he now faces the slackest possible incentive constraint.  Of course, this strategy is also inherently risky, since the expert may not possess $n$ distinct revelations, marooning the decision maker (and expert) in the region of suboptimal projects. Just how risky such a strategy is depends on the likelihood of the expert's type. 
 
In this section, I examine how this trade-off captures the general strategic concerns of the decision maker. In particular, the strategies that can be justified are exactly those that always make shortlist proposals in the `Pareto frontier' of this trade-off (i.e., no project has been revealed that is incentive compatible and would be better for the decision maker and also worse for the expert). Formally:

\begin{definition}
\label{def:locally_rat}
Call $s \in S$ \df{locally rational} if at every public history $h \in \H^{R}$, 
\begin{enumerate}[label=\textsc{\textup{(lr\arabic*)}}]
  \item\label{lr:1}  $s(h) \in \IC(h)$, and,
  \item\label{lr:2} there exists no $a \in \IC(h)$ such that $ u_d(a) \geq u_d(s(h))$ and $u_e(s(h)) \geq u_e(a)$ with at least one strict inequality.
\end{enumerate}
\end{definition}  

A locally rational strategy is one that never proposes projects which can be simultaneously improved on both fronts: increasing the decision maker's payoff should the game end, and decreasing the expert's payoff---hence the future incentive constraint---should the game continue. Locally rational strategies correspond to undominated selection rules, as defined:

\begin{definition}
\label{def:undom}
Call a monotone selection rule $f \in \F^{m}$ \df{undominated} if there is no $f' \in \F^{m}$ such that $u_d(f'(r)) \geq u_d(f(r))$ for all $r \in \Rev$ and strict for some.
\end{definition}  

We then have the following (partial) equivalence between undominated selection rules and local rationality:

\begin{theorem}
\label{thm:locally_rat_is_undominated}
Assume the environment is rich. Then
\begin{enumerate}
  \item  If $f \in F^m$ is undominated then it is implemented by a locally rational strategy $s \in S$.
  \item If $s \in S$ is locally rational then it implements an undominated $f \in \F^m$.
\end{enumerate}
\end{theorem}

Theorem \ref{thm:locally_rat_is_undominated} suggests that the only behavior that can be absolutely ruled out is the use of non locally rational strategies. On the basis of the Wald-Pearce Lemma, we can interpret Theorem \ref{thm:locally_rat_is_undominated} as stating that strategy is locally rational if and only if it is a best response to some initial probabilistic belief over the type space $\Rev$.\footnote{The Wald-Pearce Lemma (from \cite{wald1949statistical} and \cite{pearce1984rationalizable}) states that a function $g: W \to \R$ is undominated in a set $G \subseteq \R^{W}$ if and only if $g$ is an expected-value maximizing element of $G$ for some probability distribution over $G$. Of course, this abstracts away from dynamic concerns, so that while such a strategy will be an \emph{initial} best response in may not be a \emph{sequential} best response.}

Note that the converses of both statements of Theorem \ref{thm:locally_rat_is_undominated} are in general false. For (1) this is because locally rational strategies might implement multiple selection rules, some of which are dominated.\footnote{Consider for example a type space $\Rev = \{\{a\}, \{a,b\}\}$ where $u_d(a) = u_e(a) = u_e(b) = 0$ and $u_d(b) = 1$. Then the only locally rational strategy has the decision maker shortlisting $a$ initially and $b$ should it get revealed. This implements the undominated $f: \{a\} \mapsto a, \{a,b\} \mapsto b$. However, since the expert is indifferent between all projects, it also implements the dominated $f': \{a\} \mapsto a, \{a,b\} \mapsto a$.} For (2) this is because of off path behavior (i.e., there are non locally rational strategies that are outcome equivalent to locally rational ones). If we strengthen implementation to full implementation, however, the problem of multiplicity disappears, and we obtain a stronger equivalence:

\begin{corollary}
\label{cor:locally_rat_is_undominated}
Assume the environment is rich. Then a strictly monotone $f \in F^{sm}$ is undominated if and only if it is fully implemented by a locally rational strategy $s \in S$.
\end{corollary}

When inexpressibility represents unawareness or other severe epistemic constraints, it seems unreasonable for the decision maker to precisely assess the likelihood of various revelation types. When agents have limited capacity to engage with hypothetical reasoning, \cite{catonini2020local} argue for solution concepts that do not require forward planning. Greedy strategies (generalizing \emph{mostly} greedy strategies, from Section \ref{sec:eff}) do not require the decision maker to perform any forward planning. Formally:

\begin{definition}
\label{def:greedy}
Call $s \in S$ \df{greedy} if at every public history $h \in \H^{R}$, 
\begin{enumerate}[label=\textsc{\textup{(g\arabic*)}}]
  \item\label{gr:1} $s(h) \in \IC(h)$, and,
  \item\label{gr:2} for all $a \in \IC(h)$, $u_d(s(h)) \geq u_d(a)$.
\end{enumerate}
\end{definition}

Beyond their simplicity to administer, greedy strategies are compelling on the basis of the selection rules they can implement. Greedy strategies correspond to \emph{cautious} selection rules, those that, for every revelation type, maximize the worst case payoff across all more expressive types. This can be formalized as: 

\begin{definition}
\label{def:cautious}
Call a monotone selection rule $f \in \F^{m}$ \df{cautious} if there is no $f' \in \F^{m}$ and $r \in \Rev$ such that 
\begin{enumerate}
  \item\label{caut:1} $f(r') = f'(r')$ for all $r' \subsetneq r$, and,
  \item\label{caut:2} $\inf\limits_{r' \supseteq r} u_d(f'(r')) > \inf\limits_{r' \supseteq r} u_d(f(r'))$ .
\end{enumerate}
\end{definition}

Setting $r = r^\dag$ in the definition, we can see that if $f$ is cautious it maximizes the ex-ante worst case outcome. Then conditional on implementing a cautious strategy for all types (weakly) less expressive than $r$
there is no way to deviate from $f$ to another selection rule $f'$ that can guarantee a higher payoff to the decision maker than $f$ does. Specifically, if $f'$ does provide a higher worst case outcome, then it must be that $f'$ and $f$ already diverged from each other (i.e., if $f'$ satisfies \eqref{caut:2} then it must violate \eqref{caut:1}), and thus can no longer be be deviated to.

As with Theorem \ref{thm:locally_rat_is_undominated}, we have a bi-directional (but not one-to-one) correspondence between greedy strategies and cautious selection rules.

\begin{theorem}
\label{thm:greedy_is_cautious}
If $f \in F^m$ is cautious then it is implemented by a greedy strategy $s \in S$. If $s \in S$ is greedy then it implements a cautious $f \in \F^m$.
\end{theorem}

The intuition for Theorem \ref{thm:greedy_is_cautious} is straightforward: in each round, the decision maker could shortlist the same project as the prior round, and so will only propose new projects only if they increase his payoff. So, since the decision maker's payoff is increasing in the sequence of revelations, the \emph{worst} possible continuation is that the expert has nothing else to reveal; greedy optimization maximizes this lower bound.  That all cautious selection rules can be implemented as such is likewise intuitive: if $f$ deviates from the greedy optimum, then it is not maximizing the current payoff, which will be implemented should the expert have nothing further to reveal. But then the selection rule which provides the currently optimal project forever onward guarantees this lower-bound, and so $f$ could not have been cautious.

\section{Payoff Uncertainty}
\label{sec:payoff_uncertainty}

The environment set out in Sections \ref{sec:model}  does not allow for payoff uncertainty, and so, when the decision maker adds a project to the shortlist, he can be certain that the expert prefers it to any prior proposal. In this Section, I briefly examine the model in the presence of asymmetric information regarding the value of projects, in particular, when the expert (but not the decision maker) knows the realization of a payoff relevant state.

As exhibited by the Example below, uncertainty about the value of shortlisted projects can preclude normatively appealing (not to mention, focal) outcomes. These issues, however, can be circumvented with a slightly more general protocol. This generalization restores the relationship between implementability and monotonicity, albeit under and more stringent monotonicity requirement that takes into account the expert's private information about the state.

An \emph{uncertain} environment is described by a tuple $(\A,\W, (u_d,u_e), \Rev)$ where $\A$ and $\Rev$ are as before, and

\begin{itemize}
  \item $\W$ is a finite state-space
  \item $u_i : \W \times \A \to \R$ are the players'  state-dependent utility functions.
\end{itemize}

Moreover, I assume that the expert knows the state $\w \in \W$ whereas the decision maker does not. The decision maker's uncertainty can be modeled by a probability over $\W$, but it is unnecessary for the analysis that follows.

In an uncertain environment, a direct mechanism is a \emph{generalized selection rule}, a mapping from states and revelation types to feasible projects:
\begin{equation*}
g: \W \times \Rev \to \A
\end{equation*}
where $g(\w, r) \in r$ for all $\w \in \W$ and $r \in  \Rev$. Let $G$ denote the set of generalized selection rules.

To see how the addition of the expert's private information can stymie implementation, consider an environment with two states $\W = \{\w_L,\w_R\}$ and four projects
\begin{align*}
\xx   &= (\<0, 0\>, \<0, 0\>) &\qquad \bb   &= (\<2, 2\>, \<2, 2\>) \\
\aa_L &= (\<3, -1\>, \<3, -1\>) &\qquad \aa_R &= (\<-1, 3\>, \<-1, 3\>)
\end{align*}
where $z$ is described by $(\<u_d(\w_L, z), u_d(\w_R, z)\>, \<u_e(\w_L, z), u_e(\w_R, z)\>)$. Finally, let 
$$
\Rev = \big\{ r^\dag = \{\xx \}, r = \{\xx,\bb,\aa_L,\aa_R \}\big\},
$$
so that all of $\bb$, $\aa_L$, and $\aa_R$ must be revealed together. 

Initially, the decision maker is aware only of $\xx$, and so his initial shortlist is trivially $\xx$ as well. When the expert reveals (she only has one possible revelation), the decision maker must choose a new project to shortlist. Notice: if $\aa_L$ is shortlisted, then it will be rejected by (the informed) expert, should the true state be $\w_R$ and vice-versa. Hence a sufficiently risk averse decision maker will shortlist $\bb$.

This is troublesome because the two agents have \emph{completely aligned preferences}; both would strictly prefer the clearly optimal:
\begin{align*}
g (r^\dag, \w_L)   &= \xx & g (r^\dag, \w_R)   &= \xx \\
g (r, \w_L)   &= \aa_L & g (r, \w_R)   &= \aa_R 
\end{align*}
The problem is that the IRP provides no instrument by which the uninformed decision maker can delegate the choice between $\aa_L$ and $\aa_R$ to the informed expert. 

This problem can be overcome via a small generalization of the IRP, allowing the decision maker to shortlist \emph{sets} of projects at each round. That is, in the generalized IRP, the decision maker's strategy selects at each history $(r_0, p_0, \ldots, r_n)$ a subset of projects to add to the shortlist: $s(r_0, p_0, \ldots, r_n) \subseteq r_n$. In the example, after the expert reveals $r$, the decision maker could then propose $\{\aa_L,\aa_R\}$, allowing the expert the discrimination to choose the state-aligned project.

A straightforward adaptation of the Proof of Theorem \ref{thm:mondisclosure} then yields the following result.

\begin{theorem}
A generalized selection rule $g \in G$ is implemented by some strategy in the generalized IRP if and only if 
for $r, r' \in \Rev$ and $\w,\w' \in \W$
\begin{equation}
\label{eq:gen_revealing}
\tag{\textsc{m}$^*$}
 r' \subseteq r\quad \text{ implies } \quad u_e(\w, g(\w',r')) \leq u_e(\w,g(\w, r)).
\end{equation}
\end{theorem}

The incentive constraint that is required under uncertainty, \eqref{eq:gen_revealing}, is more demanding than the constraint without, \eqref{eq:fll_reavealing}, as the decision maker now has to incentivize truth telling on two separate fronts. 

Analogs of Theorems \ref{thm:full_implements}, \ref{thm:locally_rat_is_undominated}, and \ref{thm:greedy_is_cautious} all follow, although in the case of the later two, at the cost of substantially increased notational complexity. The principal issue is that the expert's strategy (i.e., order of revelations) might be conditioned on the state, thereby leading the decision maker to update his beliefs as the game unfolds. While this issue is surmountable, it adds a layer of dynamic considerations that are largely independent from the focus of this paper. Theorem \ref{thm:recvbestMech},  on the other hand, does not find an immediate counterpart in the model with payoff uncertainty. This specific failure is a manifestation of the general impossibility of facilitating ex-post efficient trade under private information, e.g., \cite{myerson1983efficient}.

The case where there is truly asymmetric information, where both the players have private information about the state, appears substantially more difficult to analyze. In this case, the expert will also update her beliefs as the game unfolds, allowing the decision maker to signal via his choice of shortlist.

\section{Discussion and Related Literature}
\label{sec:lit}
\def\L{\mathcal{L}}

The setup of this paper is closely related to the literature on project choice, \citep{armstrong2010model,guo2023regret}---and more broadly to that of delegation \citep{holmstrom1980theory,alonso2008optimal}---where an expert reveals admissible projects to a decision maker. Unlike the current model, this literature has focused on the case where the decision maker can make ex-ante commitments regarding its behavior subsequent to the expert's revelation (i.e., can commit to a direct mechanism). In contrast, I examine the case where, due to inexpressibility, unawareness, or other barriers, the decision maker cannot make such commitments.

The inability to write contracts specifying arbitrary commitments has been studied in the literature on \emph{incomplete contracts}, where, for example, \cite{grossman1986costs} consider `` a situation in which it is prohibitively difficult to think about and describe unambiguously in advance [...] all the potentially relevant aspects of the [environment].'' This literature generally takes incompleteness as given, examining the effect of this limitation on incentives and on subsequent outcomes \citep{aghion2011incomplete,hart2017incomplete}.
Rather than contemplating the direct effect of incompleteness, this paper examines how the limitations imposed by inexpressibility might be ameliorated through appropriately designed mechanisms.

The collective wisdom of the literature on unawareness and contracting---for example models of insurance \citep{filiz2012incorporating}, principal agent problems \citep{von2012incentives,auster2013asymmetric,auster2021optimal,lei2021delegation}, procurement \citep{francetich2021rationalizable}, and general contracting \citep{tirole2009cognition, piermont2017introspective}---suggests that in a wide range of contexts, asymmetric awareness leads to deliberately incomplete contracts.
This paper, while agreeing in broad strokes, offers a resolution. Even without knowing the particulars of what other agents might be aware of, the IRP yields full revelation, and thus, facilitates the completion of what would otherwise be incomplete contracts.



  \cite{maskin1999unforeseen} questioned the premise that incomplete contracts need to arise from inexpressibility, since parties could instead contract directly on \emph{utility values}, which, in specific circumstances, could themselves be revealed in an incentive compatible way. Their argument critically rests on the specifics of the environment they considered, chiefly, transferable utility and the ability ``to perform dynamic programming [and thus] formulate a probability distribution over the possible payoffs." This paper offers a different process of circumventing incomplete contracts that eschews both of these assumptions.

There is a very small set of papers that directly examine the problem of eliciting awareness. First, \cite{herweg2020procurement} consider a procurement problem in which the seller may be aware of some design flaws not conceived of by the buyer. The authors derive an efficient mechanism using competition between sellers to incentivize revelation.

Second, and more closely related, \cite{pram2023efficient} consider the effects of unawareness in a general mechanism design environment with transferable (i.e., quasi-linear) utility. Specifically, they introduce \emph{dynamic elaboration VCG mechanisms} in which agents report payoff types, the mediator communicates back the pooled awareness, and agents can further elaborate on their reported types. This is repeated till nobody wants to change their reports after which a outcome is enacted using a VCG mechanism plus some additional term incentivizing agents to raise awareness. Despite the essential differences between their model and the present one, the insights from both papers roughly agree: efficient and fully revealing mechanisms will entertain a dynamic component whereby agents can reveal their awareness in piecemeal fashion. Moreover, both papers stress the importance of commitment, not to specific outcomes, but to the procedure by which outcomes are selected.

Finally, \cite{hanany2025robust} consider a robust contracting environment where the principal allows the agent to voluntarily disclose novel actions. In contrast to the case without disclosure, when the agent can disclose novel actions, the optimal robust contract may be non-linear. To abstract away from dynamic contracting concerns, the authors examine the case where the agent is aware of at most one additional action.  


In formally considering what agents can express, this paper is related to syntactic approaches to economics that directly model the language used by agents. For example in equilibrium selection \citep{demichelis2008language}, contracting \citep{jakobsen2020model,piermont2023vague}, psychological framing \citep{kahneman1984choices}, translation \citep{piermont2025do}, and the general theory of games \citep{bjorndahl2013language} and decisions \citep{blume2021constructive,piermont2024failures}.

When an agent reveals some set of projects $r$, this revelation is \emph{evidentiary}: it is costless to conceal and verifiably true. As such, this paper is related to the literature on mechanism design with evidence, where one party can (voluntarily) prove something about the state \citep{dye1985disclosure,green1986partially,grossman1986costs,bull2007hard,ben2019mechanisms}. In contrast to these papers, here the evidence is not simply informational but regards the existence of \emph{actions}, directly altering the choice set of the decision maker. Despite this difference, the results in this paper broadly reflect the insight that full ex-ante commitment is often unnecessary, since by designing the appropriate, game agents will voluntarily reveal evidence whenever it is valuable.

The present paper is also related to the literature on robust mechanism design, which seeks to find mechanisms that implement desirable outcomes in the absence of stringent common knowledge assumptions. Often, this means the construction of mechanisms agnostic to agents' types \citep{bergemann2005robust,jehiel2006limits}. In the IRP, greedy strategies are robust in this sense as they do not rely on reasoning about the expert's type and instead provide a worst-case guarantee across all possible types. For a non-technical overview of some of the recent contributions to robust contract theory and mechanism design, see \cite{carroll2019robustness}.

\appendix


\section{Proofs}

\begin{tproof}{thm:mondisclosure}
(\ref{md:ic} $\implies$ \ref{md:*}) Let $f \in \F$ be monotone, and define $s^\dag \in S$ via
$$s^\dag(r_0, p_0, \ldots, r_n) =  f(r_n).$$
 Fix $r \in \Rev$ and let $\ss^{r}$ be the strategy which reveals $r$ at the first opportunity, and then chooses $f(r)$.
Then, let $\ss' \in \Sigma(r)$ be any other strategy. Since the expert can never reveal something more expressive than $r$, it follows that $a(s^\dag,\ss') = f(r')$ some $r' \subseteq r$. Then,
\begin{align*}
  u_e(a(s^\dag,\ss^{r})) &= u_e(f(r)) && \text{(by def. of $s^\dag$, $\ss^{r}$)}\\
  &\geq u_e(f(r')) && \text{(by monotonicity)}\\
  &= u_e(a(s^\dag,\ss')) && \text{(by def. of $r'$)}
\end{align*}
Thus, $\ss^{r}  \in \BRR(s^\dag,r)$: $s^\dag$ transparently implements $f$.

(\ref{md:*} $\implies$ \ref{md:im}) Follows immediately from the containment $\BRR(s,r) \subseteq \BR(s,r)$.

(\ref{md:im} $\implies$ \ref{md:ic}) Let $f \in \F$ be implemented by $s \in S$. Fix some $r,r' \in \Rev$ with $r' \subseteq r$. By our assumption, there exists some $\ss \in \BR(s,r)$ such that $f(r) = a(s,\ss)$ and $\ss' \in \BR(s,r')$ such that $f(r') = a(s,\ss')$. Notice, $\ss' \in \Sigma(r') \subseteq \Sigma(r)$. It follows that
$$
u_e(f(r)) = \max_{\ss \in \Sigma(r) } u_e(a(s,\ss)) \geq u_e(a(s, \ss'))  = u_e(f(r'))
$$
so $f$ is monotone.
\end{tproof}

\begin{tproof}{thm:full_implements}
($\rightarrow$) Let $f \in \F$ be strictly monotone. Fix $r \in \Rev$. Define $s^\dag$ and $\ss^{r}$ as in Part (\ref{md:ic} $\implies$ \ref{md:*}) of proof of Theorem \ref{thm:mondisclosure}.  By that proof $\ss^{r} \in \BR(s^\dag, r)$. Now, let $\ss' \in \Sigma(r)$ be such that $a(s^\dag,\ss') \neq f(r)$; it follows that $a(s^\dag,\ss') = f(r')$ for some $r' \subseteq r$.
Then: 
\begin{align*}
  u_e(a(s^\dag,\ss^{r})) &= u_e(f(r)) && \text{(by def. of $s^\dag$, $\ss^{r}$)}\\
  &> u_e(f(r')) && \text{(by strict monotonicity)}\\
  &=u_e(a(s^\dag,\ss')) && \text{(by def. of $r'$)}
\end{align*}
Thus, $\ss' \notin \BR(s,r)$. It follows that $s^\dag$ fully implements $f$.

($\leftarrow$) Let $s \in S$ fully implement $f \in \F$.
By Theorem \ref{thm:mondisclosure}, $f$ is monotone. 
Fix $r,r' \in \Rev$ with $r' \subseteq r$ such that $f(r') \neq f(r)$. Let $\ss \in \BR(s,r)$, $\ss' \in \BR(s,r')$; by definition of full implementation, $a(s,\ss) = f(r)$ and $a(s,\ss') = f(r')$, and by its contra-positive, $\ss' \notin \BR(s,r)$. We have $u_e(f(r)) = u_e(a(s,\ss)) > u_e(a(s,\ss')) = u_e(f(r'))$; $f$ is strictly monotone.
\end{tproof}

\begin{tproof}{thm:recvbestMech}
Let $s \in S$ be mostly greedy as defined in Definition \ref{def:mostly_greedy}.
Let $\rhd$ be a total ordering of $\A$ that extends $u_d(\cdot)$. That is $\rhd$ is an anti-symmetric weak order such that $u_d(a) \geq u_d(a')$ implies $a \rhd a'$.\footnote{Intuitively, $\rhd$ represents some tie-breaking rule to compare those projects that yield equal utility to the decision maker.} 

For each $r \in \Rev$, let $\ss^{r} \in \BR(s,r)$ be a strategy that yields the $\rhd$-maximal project across all best responses to $s$. Note that $a(s, \ss^r)$ is uniquely determined, owing to the fact that $\rhd$ is a total order. Let $f: r \mapsto a(s, \ss^r)$. 

Clearly $s$ implements $f$; we will show that $f$ is efficient. Fix some $r \in \Rev$. Set $a^\star = a(s, \ss^r)$. There must exist some public history $h = (r_0, p_0, \ldots, r_n)$ with $r_n \subseteq r$ such that $s(h) = a^\star$. 

If $r_n \neq r$, consider the extended public history $h' = (r_0, p_0, \ldots, r_n, p_n, r)$; set $a' = s(h')$. Notice that $u_e(a') \geq u_e(a^{\star})$ since $a' \in \IC(h')$ and $u_e(a') \leq u_e(a^{\star})$ since $\ss^r$ is a best response. Likewise, notice that  $u_d(a') \geq u_d(a^{\star})$ by \scref{mgr:2} and $u_d(a') \leq u_d(a^{\star})$ since $\ss^r$ maximizes $\rhd$. 

Take some $a'' \in r$, there are two cases: if  $a'' \in r \setminus \IC(h')$, then $u_e(a^{\star}) = u_e(a') > u_e(a'')$, so $a''$ does not Pareto dominate $f(r) = a^{\star}$.
Alternatively if $a'' \in r \cap \IC(h')$, then by \scref{mgr:2}, $u_d(a^{\star}) = u_d(a') \geq u_d(a'')$; moreover, if this holds with equality then by \scref{mgr:3}, $u_e(a^{\star}) = u_e(a') \geq u_e(a'')$. Again $a''$ does not Pareto dominate $f(r)$. So, $f$ is efficient.

Now assume that $f'$ is any other monotone and efficient selection rule.
By way of contradiction, assume that for some $r \in \Rev$, we have $u_d(f'(r)) > u_d(f(r))$. Without loss of generality, take this to be a `minimal' such occurrence: that is, that for all $r' \subsetneq r$ we have $u_d(f(r')) \geq u_d(f'(r'))$. By the efficiency of $f'$ this implies
\begin{equation}
\label{eq:poimp}
u_e(f'(r')) \geq u_e(f(r'))
\end{equation}
for all such $r'$.

Let $h = (r_0, p_0, \ldots, r_n)$ be a public history such that $r_n \subseteq r$ and $s(h) = a(s, \ss^r)$. Let $a' \in \bigcup_{i < n}\{p_i\}$. We have
\begin{align*}
u_e(f'(r)) 
&\geq u_e(f'(r_n)) && \text{(since $f'$ is monotone)}\\
&\geq u_e(f(r_n)) &&\text{(by \eqref{eq:poimp})}\\
& \geq u_e(a') &&\text{(since $\ss^r \in \BR(s,r)$)}
\end{align*}
Thus, $f'(r) \in \IC(h)$. So by \scref{mgr:2}, we must have $u_d(f(r)) \geq u_d(f'(r))$, contradicting our assumption. It holds that no such $r$ exists.
\end{tproof}

\begin{tproof}{thm:compStatics}
 Let $s^{\Rev}$ and $s^{\Qev}$ denote the mostly greedy strategies for type spaces $\Rev$ and $\Qev$. 

\begin{enumerate}
  \item Fix $r \in \Rev$ and let $\Qev = \{r' \in \Rev \mid r \subseteq r'\}$. We will show by induction that for all $q \in \Rev \cap \Qev$, 
\begin{align}
u_e(f^{\Rev}(q)) &\geq u_e(f^{\Qev}(q)), \text{ and } \label{eq:inductiveclaimproofhereXX} \\
u_d(f^{\Rev}(q)) &\leq u_d(f^{\Qev}(q)) \label{eq:inductiveclaimproofhere}
\end{align}
with \eqref{eq:inductiveclaimproofhere} yielding the claim.

The base cases is $q = q^\dag =  r$. Let $h = (r)$ be the initial history for type space $\Qev$. Notice that $\IC(h) = r$. So $f^{\Rev}(r) \in \IC(h)$; if follows that
$$
u_d(f^{\Qev}(r)) = u_d(s^{q}(h)) \geq u_d(f^{\Rev}(r)),
$$
where the inequality arises from \scref{mgr:2}.
This establishes \eqref{eq:inductiveclaimproofhere}.

Assume $u_e(f^{\Rev}(r)) \leq u_e(f^{\Qev}(r))$. Now let $h'$ be a history (in the environment with type space $\Rev$) ending in $r$ that results from $s^\Rev$ and the expert's best response to it that implements $f^\Rev$: it follows that $f^{\Qev}(r) \in \IC(h')$: Thus by \scref{mgr:2} of mostly greedy, $u_d(f^{\Rev}(r)) \geq u_d(f^{\Qev}(r))$, and hence, by \eqref{eq:inductiveclaimproofhere}, as just established above, $u_d(f^{\Rev}(r)) = u_d(f^{\Qev}(r))$. Then by \scref{mgr:3} we have $u_e(f^{\Rev}(r)) \geq u_e(f^{\Qev}(r))$, which yields \eqref{eq:inductiveclaimproofhereXX}.

The inductive step is almost identical, noting that at histories ending in the same revelation, $s^{\Rev}$ maximizes $u_d$ relative to a weaker constraint than $s^{\Qev}$ (this is the inductive hypothesis) and so establishes \eqref{eq:inductiveclaimproofhere}. Given this, \eqref{eq:inductiveclaimproofhereXX} follows from an identical argument to the base case.

\item As the Example from Section \ref{sec:ex}, let 
$$\A = \{\yy = (0,0),  \xx = (2,2), \aa = (3,3), \bb = (4,1)\}
$$
and 
$$
\Rev = \big\{ \{\yy\}, \{\yy,\xx\}, \{\yy, \aa,\bb\}, \{\yy, \xx, \aa,\bb\}\big\}.
$$
So that $f^{\Rev}$ is defined by
$$
\{\yy\} \mapsto \yy, \quad \{\yy,\xx\} \mapsto \xx, \quad \{\yy, \aa,\bb\} \mapsto \bb,  \quad \{\yy, \xx, \aa,\bb\}  \mapsto \aa
$$
We have $u_d(f^{\Rev}(\{\yy, \aa,\bb\})) > u_d(f^{\Rev}(\{\yy, \xx, \aa,\bb\})) > u_d(f^{\Rev}(\{\yy\}))$.

\item Notice that the mostly greedy strategy depends only on what has been revealed up to that point. As such, if $(r_0, p_0, \ldots, r_n)$ and $(q_0, p_0, \ldots, q_n)$ are public histories such that $r_i = q_i$ for $i \leq n$, then $s^{\Rev}(r_0, p_0, \ldots, r_n) = s^{\Qev}(q_0, p_0, \ldots, q_n)$. It follows that any project that can be chosen (via some strategy) by revelation type $r \in \Rev \cap \Qev$ when the type space is $\Qev$ can also be chosen when the type space is $\Rev$ (by mimicking the same strategy). As such it must be that $u_e(f^{\Rev}(r)) \geq u_e(f^{\Qev}(r))$. Efficiency then requires that $u_d(f^{\Rev}(q)) \leq u_d(f^{\Qev}(q))$.

\item Let $\Rev = 2^\A$ and fix $r \in \Rev$. Set 
$$
a^\star_r \in \argmax \{ u_e(a) \mid a \text{ is Pareto efficient in } r, u_d(a) \geq \max_{a' \in r^\dag} u_d(a') \}
$$ as an expert's most preferred efficient project from $r$ that is at least as good as for the decision maker as what he could achieve alone. If there are multiple such projects, note that they must be payoff equivalent; so assume without loss of generality that it is unique. Notice that efficiency / individual rationality requires that 
$$f^{\Rev}(r^\dag \cup \{a^\star_r\}) = a^\star_r.$$
Thus, by monotonicity $u_e(f^{\Rev}(r)) \geq u_e(a^\star_r)$; but since $a^\star_r$ yields the highest possible payoff to the expert's over all efficient outcomes in $r$, it must be that $f(r) = a^\star_r$.
\end{enumerate}
\end{tproof}

\begin{tproof}{thm:locally_rat_is_undominated}
($1$) Let $f \in \F^{m}$ be undominated. Construct the strategy $s \in S$ inductively on the length of the history. Set $s(r^\dag) = f(r^\dag)$ and then for each $h = (r_0, p_0, \ldots, r_n)$ let
\begin{itemize}
  \item $s(h) = f(r_n)$ if there is no $a \in \IC(h)$ such that $ u_d(a) \geq u_d(f(r_n))$ and $u_e(f(r_n)) \geq u_e(a)$ with at least one strict inequality, and,
  \item $s(h) \in \argmax \{u_d(a) \mid a \in \argmin\{u_e(a') \mid a' \in \IC(h)\}\}$ otherwise.
\end{itemize}
The strategy $s$ is locally rational by construction, so the claim holds as long as $s$ implements $f$. 
We will show this by induction on $r$. For $r = r^\dag$ it is by definition, so assume that for all $r' \subsetneq r$, there is some $h'$ such that $s(h') = f(r')$, such that $h'$ arises as a best response for type $r'$ with the expert ultimately choosing to enact $s(h')$.

Let $h$ result from $r$ playing a best response to $s$. Without loss of generality assume $h$ ends in full revelation of $r$. 
Assume by way of contradiction that $s(h) \neq f(r)$. Notice that by the inductive hypothesis, the payoff of every type $r' \subsetneq r$ us $u_e(f(r'))$; thus by the monotonicity of $f$, $f(r) \in \IC(h)$.
As such, there must exist some $a \in \IC(h)$ such that $u_d(a) \geq u_d(f(r))$ and $u_e(f(r)) \geq u_e(a)$ with at least one strict inequality. There are two cases:
\begin{itemize}
  \item $u_d(a) > u_d(f(r))$: Then, define $f'$ via $f'(r^{\diamond}) = f(r^{\diamond})$ for all $r^{\diamond}\neq r$ and $f(r) = a$. Clearly $f'$ is monotone and dominates $f$, a contradiction.
  \item If $u_d(a) = u_d(f(r))$ and $u_e(f(r)) > u_e(a)$.  Let $r^{\circ} = r \cup \{a^{\circ}\}$ for some project $a^{\circ} \in \A$ such that  
  $$
  u_d(a^\circ) > \max_{a \in r} u_d(a) \qquad \text{ and } \qquad  u_e(f(r)) > u_e(a^\circ) > u_e(a),
  $$
  which exists by the richness assumption. Then $f'$ defined by $f'(r^{\diamond}) = f(r^{\diamond})$ for all $r^{\diamond}\notin \{r, r ^\circ\}$ and $f'(r) = a$ and $f'(r^\circ) = a^\circ$. Again $f'$ is monotone and dominates $f$, (note that $f(r^\circ) \neq a^\circ$ by monotonicity).
\end{itemize}
Thus, it must be that for all $r$, there exists some history $h$, induced by a best response, such that $s(h) = f(r)$. Since $s(h) \in \IC(h)$, it is a best response for the expert to choose $s(h)$.

($2$) Let $s$ be locally rational. As in the proof of Theorem \ref{thm:recvbestMech}, let $\rhd$ be a total ordering of $\A$ that extends $u_d(\cdot)$.  For each $r \in \R$ let $\ss^{r} \in \BR(s,r)$ be a strategy that yields the $\rhd$-maximal project across all best responses to $s$. Note that $a(s, \ss^r)$ is uniquely determined, owing to the fact that $\rhd$ is a total order. Let $f: r \mapsto a(s, \ss^r)$. 

Now let $f'$ denote any other monotone selection rule. We will show that $f'$ does not dominate $f$. Assume that there exists some $r$ such that $u_i(f'(r)) \neq u_i(f(r))$ for some $i \in \{d,e\}$, (if there is not such $r$, then the claim holds trivially). Without loss of generality, assume this is a minimal instance, so that for all $r' \subsetneq r$, we have $u_i(f(r')) = u_i(f'(r'))$.

If $u_d(f(r)) > u_d(f'(r))$ then $f'$ does not dominate $f$, so assume that $u_d(f'(r)) \geq u_d(f(r))$. Now, let $h$ result from $r$ playing $\ss^r$. Note two things: 

\begin{enumerate}[label=(\roman*)]
  \item $f'(r) \in \IC(h)$: this follows from the monotonicity of $f'$, and the fact that we have assumed that for all $r' \subsetneq r$, $u_e(f'(r')) = u_e(f(r'))$.
  \item $u_i(f(r)) = u_i(s(h))$ for $i \in \{d,e\}$: since $f(r)$ results from a best response to $s$ it follows that $u_e(f(r)) \geq u_e(s(h))$ and since $s(h) \in \IC(h)$, we have $u_e(s(h)) \geq u_e(f(r))$. Then, since the strategy $\ss^r$ chooses the $\rhd$-preferred best response, it follows that $u_d(f(r)) \geq u_d(s(h))$. Since $s$ is locally rational, and since $u_e(s(h)) = u_e(f(r))$, it follows from  \scref{lr:2} that $u_d(s(h)) \geq u_d(f(r))$. Note: it is possible that $s(h) \neq f(r)$ if both yield the same material payoffs to both players but $f(r)$ is $\rhd$-preferred to $s(h)$.
\end{enumerate}

Combining (i) and (ii), the locally rationality of $s$, i.e., \scref{lr:2}, requires that $u_e(f'(r)) > u_e(s(h)) = u_e(f(r))$. Let $r^{\circ} = r \cup \{a^{\circ}\}$ for some project $a^{\circ} \in \A$ such that  
  $$
  u_d(a^\circ) > \max_{a \in r} u_d(a) \qquad \text{ and } \qquad  u_e(f'(r)) > u_e(a^\circ) > u_e(f(r)),
  $$
  which exists by the richness assumption. Then $s(h,r^\circ) = a^\circ$ by local rationality so $f(r^\circ) = a^\circ$. So $f'$ does not dominate $f$, (note that $f'(r^\circ) \neq a^\circ$ by monotonicity).
\end{tproof}

\begin{tproof}{thm:greedy_is_cautious}
($1$) Let $f \in \F^{m}$ be cautious. Construct the strategy $s \in S$ inductively on the length of the history. Set $s(r^\dag) = f(r^\dag)$ and then for each $h = (r_0, p_0, \ldots, r_n)$ let
\begin{itemize}
  \item $s(h) = f(r_n)$ if $f(r_n) \in \IC(h)$ and there is no $a \in \IC(h)$ such that $ u_d(a) > u_d(f(r_n))$, and,
  \item $s(h) \in \argmax \{u_d(a) \mid a \in \IC(h)\}\}$ otherwise.
\end{itemize}

The strategy $s$ is greedy by construction, so the claim holds as long as $s$ implements $f$. 
We will show this by induction on $r$. For $r = r^\dag$ it is by definition, so assume that for all $r' \subsetneq r$, there is some $h'$ such that $s(h') = f(r')$, such that $h'$ arises as a best response for type $r'$ with the expert ultimately choosing to enact $s(h')$.

Let $h$ result from $r$ playing a best response to $s$. Without loss of generality assume $h$ ends in full revelation of $r$. 
Assume by way of contradiction that $s(h) \neq f(r)$. Notice that by the inductive hypothesis, the payoff of every type $r' \subsetneq r$ us $u_e(f(r'))$; thus by the monotonicity of $f$, $f(r) \in \IC(h)$. As such, there must exist some $a \in \IC(h)$ such that $u_d(a) > u_d(f(r))$.

But then, consider $f'$ defined by $f'(r^{\diamond}) = a$ for all $r^{\diamond} \supseteq r$ and $f'(r') = f(r')$ otherwise. Then, $f'$ is monotone and  
$$\inf\limits_{r' \supseteq r} u_d(f'(r')) = u_d(a) > u_d(f(r))  > \inf\limits_{r' \supseteq r} u_d(f(r')),$$
 contradicting the cautiousness of $f$.

($2$) Let $s$ be greedy.  As in the proof of Theorem \ref{thm:recvbestMech}, let $\rhd$ be a total ordering of $\A$ that extends $u_d(\cdot)$.  For each $r \in \R$ let $\ss^{r} \in \BR(s,r)$ be a strategy that yields the $\rhd$-maximal project across all best responses to $s$. Note that $a(s, \ss^r)$ is uniquely determined, owing to the fact that $\rhd$ is a total order. Let $f: r \mapsto a(s, \ss^r)$. 

First we claim: if $r' \supseteq r$ then $u_d(f(r')) \geq u_d(f(r))$. For a fixed $r$, we can show this by induction on $r'$, assume it holds for all $r''$ such that $r \subseteq r'' \subsetneq r'$. Let $h = (r_0, p_0, \ldots, r_{n-1}, p_{n-1}, r_n)$ be the result of playing $\ss^{r'}$ such that $s(h) = f(r')$. Note, since $f(r_{n-1}) = a(s, \ss^{r_{n-1}})$ arises from a best response, it follows that 
$u_e(f(r_{n-1})) \in \IC(h)$. Thus, by the greediness of $s$, we have that $u_d(f(r')) = u_d(s(h)) \geq u_d(f(r_{n-1})) \geq u_d(f(r))$, where the last inequality is the inductive hypothesis.

Now, fix $r$ and take any other monotone $f' \in F^m$ such that $f(r') = f'(r')$ for all $r' \subsetneq r$.  
Let $h$ be the result of playing $\ss^{r}$ such that $s(h) = f(r)$. 
From the monotonicity of $f'$, it follows that $f'(r) \in \IC(h)$, and hence by the greediness of $s$:
$$
\inf\limits_{r' \supseteq r} u_d(f(r')) = f(r) = s(h) \geq f'(r) \geq \inf\limits_{r' \supseteq r}u_d(f'(r')). 
$$
So $f$ is cautious.
\end{tproof}

\newpage
\singlespacing
\small

\setlength{\bibsep}{1pt plus 0.3ex}

\bibliographystyle{aer}
\bibliography{EA.bib}

\end{document}